\documentclass{aa}  
\usepackage[table]{xcolor}

\makeatother
\usepackage{graphicx}
\usepackage{txfonts}
\usepackage[colorlinks=true,linkcolor=blue,citecolor=blue]{hyperref}
\usepackage{url}
\usepackage{mathtools}
 \usepackage{booktabs}
 \usepackage{placeins}
\begin{document}

   \title{Decoding the IRX-$\beta$ dust attenuation relation in star-forming galaxies at intermediate redshift 
}

\author{M. Hamed\inst{1}
          \and
           F. Pistis\inst{1}
           \and
           M. Figueira\inst{1,4}
           \and
           K. Ma\l{}ek\inst{1,2}
           \and
                  A. Nanni\inst{1,5}
           \and
            V. Buat\inst{2}
            \and
           A. Pollo\inst{1,3}
           \and
           D. Vergani\inst{6}
          \and
          M. Bolzonella\inst{6}
          \and
          Junais\inst{1}
          \and
          J. Krywult\inst{7}
          \and
          T. Takeuchi\inst{8}
          \and
         G. Riccio\inst{1}
         \and
         T. Moutard\inst{2}
          }

         \institute{National Centre for Nuclear Research, ul. Pasteura 7, 02-093 Warszawa, Poland
         \and
         Aix Marseille Univ. CNRS, CNES, LAM, Marseille, France
         \and
         Astronomical Observatory of the Jagiellonian University, ul.Orla 171, 30-244 Krakow, Poland
         \and
         Institute of Astronomy, Faculty of Physics, Astronomy and Informatics, UMK, ul. Grudziądzka 5, 87-100 Toruń, Poland
         \and
         INAF - Osservatorio astronomico d'Abruzzo, Via Maggini SNC 64100 Teramo, Italy
         \and
         INAF - Osservatorio di Astrofisica e Scienza dello Spazio di Bologna, Via Piero Gobetti 93/3, I-40129 Bologna, Italy
         \and
         Institute of Physics, Jan Kochanowski Univesity, ul. Swietokrzyska 15, 25-406 Kielce, Poland
         \and
         Division of Particle and Astrophysical Science, Nagoya University, Furo-cho, Chikusa-ku, Nagoya, Aichi 464-8602, Japan
         }

   \date{}

 
  \abstract
  {}
  {We aim to understand what drives the IRX-$\beta$ dust attenuation relation at intermediate redshift ($0.5 < z < 0.8$) in star-forming galaxies. We investigate the role of various galaxy properties in shaping this observed relation.} 
  {We use robust $\left[ \ion{O}{ii} \right] \lambda 3727$,  $\left[ \ion{O}{iii} \right] \lambda\lambda 4959, 5007$, and H$\beta$ line detections of our statistical sample of 1049 galaxies to estimate the gas-phase metallicities. We derive key physical properties that are necessary to study galaxy evolution, such as the stellar masses and the star formation rates, using the spectral energy distribution fitting tool \texttt{CIGALE}. Equivalently, we study the effect of galaxy morphology (mainly the S\'ersic index $n$ and galaxy inclination) on the observed IRX-$\beta$ scatter. We also investigate the role of the environment in shaping dust attenuation in our sample.}
  {We find a strong correlation of the IRX-$\beta$ relation on gas-phase metallicity in our sample, and also strong correlation with galaxy compactness characterized by the S\'ersic indexes. With higher metallicities, galaxies move along the track of the IRX-$\beta$ relation towards higher IRX. Correlations are also seen with stellar masses, specific star formation rates and the stellar ages of our sources. Metallicity strongly correlates with the  IRX-$\beta$ scatter, this also results from the older stars and higher masses at higher beta values. Galaxies with higher metallicities show higher IRX and higher beta values. The correlation with specific dust mass strongly shifts the galaxies away from the IRX-$\beta$ relation towards lower $\beta$ values. We find that more compact galaxies witness a larger amount of attenuation than less compact galaxies. There is a subtle variation in the dust attenuation scatter between edge-on and face-on galaxies, but the difference is not statistically significant. Galaxy environments do not significantly affect dust attenuation in our sample of star-forming galaxies at intermediate redshift.}
   {}


   \maketitle
%

\section{Introduction}

Extragalactic astronomy witnessed major development in the last few decades with regard to observation and understanding of the physical and chemical processes that control the evolution of galaxies. This understanding was a result of interpreting the ever-growing plethora of panchromatic data, that provided unprecedented constraints on the interplay between the different components that galaxies exhibit. With powerful infrared (IR) telescopes such as the \emph{Herschel} Space Observatory, it has become possible to better constrain the direct and indirect stellar emission of galaxies at higher redshifts, expanding our knowledge beyond the highly-resolved local Universe.\\

In complex systems like galaxies, different components interact with each other on different timescales. Such interaction includes the dust attenuation of the stellar light, which influences the total spectra of galaxies. Dust affects the shape of the spectral energy distribution (SED) like no other component, despite its low contribution to the overall mass of the baryonic matter. Interstellar dust absorbs a significant amount of ultraviolet (UV) and optical radiation, heats up, and re-emits it at longer wavelengths, mostly in the far-infrared (FIR). As a consequence, part of the stellar UV emission gets extinct, making it necessary to account for the missing radiation especially when estimating key properties that describe the evolution of galaxies, such as star formation rates (SFRs). The derivation of SFR should rely on both UV-optical measurements and on the FIR emission \citep[e.g.,][]{Blain2002,Takeuchi2005,Chapman2005,Hopkins06,Madau14,Magnelli2014,Bourne2017,Whitaker2017,Gruppioni2020}. Dust attenuation laws are used to correct the absorption of the short wavelength photons in order to recover fundamental properties of galaxies. This is typically done by assuming certain dust distribution relative to the dimmed stellar populations. Attenuation laws rely on the well studied dust extinction in nearby galaxies \citep[e.g.,][]{Calzetti1994, Calzetti2000, CharlotFall2000, Johnson2007}, and they are widely applied when reproducing the UV-optical spectra at different redshifts \citep[for an extensive review on attenuation laws see][]{Salim2020}.\\

At higher redshifts, the challenging measurements of FIR emission are overpowered by the easily available rest-frame UV emission \citep[e.g.,][]{Burgarella2007, daddi2007, Bouwens2012}. This in turn limits the wavelength range from which the physical properties are inferred; therefore, a correct understanding of physical processes that prevail in the short wavelength domain, like dust attenuation, becomes critical. \citet[][]{Calzetti1994} showed a correlation between the UV spectral slope $\beta$, which is indicative of attenuation, measured with different spectral windows, and the Balmer optical depth of local starburst galaxies. Using the same sample, \citet[][]{Meurer1995,Meurer99} found a tight relation between the heavily-attenuated $\beta$ and the IR excess (IRX) of galaxies defined as the ratio between the IR and FUV luminosities log(L$_{\text{IR}}$/L$_{\text{FUV}}$). This became known as the IRX-$\beta$ relation, and it was well observed subsequently in numerous studies at low and high redshift \citep[e.g.,][]{Overzier2011, Boquien2012, Takeuchi2012, Buat12, Cullen2017, Calzetti2021, Schouws2021}.

Ideally, such a relation can be used to infer the FIR luminosity and affiliated properties of galaxies from lone UV observations. However, outliers of the IRX-$\beta$ relation were found in several samples at different redshift ranges \citep[e.g.,][]{Casey2014, Alvarez2016,McLure2018}. These outliers are typically Ultra Luminous IR Galaxies (ULIRGs) and populate the region of higher IRX and lower $\beta$ values. Moreover, in older works, $\beta$ tends to be biased towards lower values due to the underestimated UV fluxes from the small aperture of \emph{International Ultraviolet Explorer} (IUE), and this relation was therefore revisited using the fluxes from the \emph{Galaxy Evolution Explorer} (GALEX) \citep[][]{Takeuchi2010,Takeuchi2012}, moving it to higher $\beta$. Furthermore, the interpretation of this relation is not fully understood, despite the numerous attempts that tried to unveil the factors upon which it relies. The attenuation curve used to reproduce the observed short wavelengths, as well as the dust geometry model used to attenuate, were found to strongly affect the IRX-$\beta$ scatter \citep[][]{Boquien2009,Casey2014,Salmon2016}. This non-universality of dust attenuation laws is a known feature of dust obscuration at different redshift ranges \citep[e.g.,][]{Kriek2013, Buat2018, buat2019, Hamed21}. Other dependencies of the IRX-$\beta$ relation are the age of stellar populations \citep[][]{Popping2017, Reddy2018}, the molecular gas mass \citep[][]{Ferrara2017}, and on gas-phase metallicity \citep[][]{Reddy2018,Shivaei2020}.

The IRX-$\beta$ relation was well studied in the local Universe and at high redshift. In this work, we aim to answer the key question of what drives the IRX-$\beta$ relation at intermediate redshift for star-forming galaxies, putting the evolution of such relation in the context of galaxy evolution. We make use of robust metallicity estimations, galaxy morphological quantities, and other physical properties, such as stellar mass, in order to understand the attenuation scatter and therefore better understand this relation at intermediate redshift, an area that otherwise remains poorly studied.

In this work, we use the definition of the $IRX\, =\, log(L_{IR}/L_{FUV})$, and that of $\beta = \frac{log(F_{NUV}/F_{FUV})}{log(\lambda_{FUV}/\lambda_{NUV})}-2$. This paper is structured as follows: in Section \ref{data}, we describe the data used in this work. In section \ref{sed} we discuss the SED fitting procedure and the derivation of key physical properties that govern galaxy evolution. In section \ref{mock} we discuss the quality of our SED fitting procedure. In section \ref{irx_beta}, we show the IRX-$\beta$ relation of our sample with its fit. We derive the gas-phase metallicities of our sample and discuss its effect on the IRX-$\beta$ scatter in section \ref{metal}. In section \ref{discussion}, we discuss different physical properties that drive the IRX-$\beta$ scatter, with the dependence on the galaxy environment in section \ref{env}. The conclusions are presented in section \ref{conclusions}. Throughout this paper, we adopt the stellar IMF of \citet{Chabrier2003} and a $\Lambda$CDM cosmology parameters (WMAP7, \citealp{Komatsu2011}): H$_0$ = 70.4 km s$^{-1}$ Mpc$^{-1}$, $\Omega_{M}$ = 0.272, and $\Omega_{\Lambda}$ = 0.728.

\section{Data}\label{data}

\subsection{Spectroscopic data}\label{spectro}
The data used in this work are from the VIMOS Public Extragalactic Redshift Survey \citep[VIPERS,][]{Guzzo2014, Garilli2014, Scodeggio2018}. VIPERS used the VIMOS spectrograph at the Very Large Telescope to measure redshifts for a 
large number of galaxies ($\sim$90\,000) in 
{\bf two fields covering} a total of $\sim24\ \text{deg}^2$ at $0.5\le z\le 1.2$. The VIPERS sample has a limiting magnitude of $i_{AB} \leq 22.5$ mag to maximize the signal-to-noise of the spectra and to select galaxies below $z=1.2$ \citep{Guzzo2014}.  
Moreover, a color selection based on the \textit{u,g,r,i} bands allowed removing galaxies below $z=0.5$, with median redshift 0.7 \citep{Guzzo2014}.\\
Spectroscopic observations of VIPERS were collected using the low-resolution red (LR-Red) grism ($\sim$ 5500-9500\AA) and a spectral resolution R $\sim$ 220 \citep{Scodeggio2018}. 
VIPERS data reduction was performed via a fully automated pipeline \citep{Garilli2014}. Redshifts were estimated using the EZ code \citep{Garilli2010} aided by the visual inspection. The redshift quality flag, described in detail in \citealp{Garilli2014} and \citealp{Guzzo2014}, has tentative (flag 1) to  secure (flags 2 to 4, with at least 90\% of confidence) redshift measurements.

For reliable redshift estimations \citep{Scodeggio2018}, we selected galaxies for which the redshift flag was $3.0 \leq z_{flag} \leq 4.5$, corresponding to redshift confidence $> 99\%$. 
We also selected the VIPERS sample with a flag that provides measurements with the quality of the measurement of emission lines. This flag is a goodness of fit assessment \citep[a thorough description is detailed in][]{Pistis2022, Figueira2022}. The flags of emission lines are well-checked to deliver good estimations of the line profile. This is done through certain criteria such as minimizing the distance between the peak of the fit and the brightest pixel, and constraining the fit amplitudes to not differ significantly from the observed emission \citep{Pistis2022}.

\begin{figure}
    \centering
    \includegraphics[width=\columnwidth]{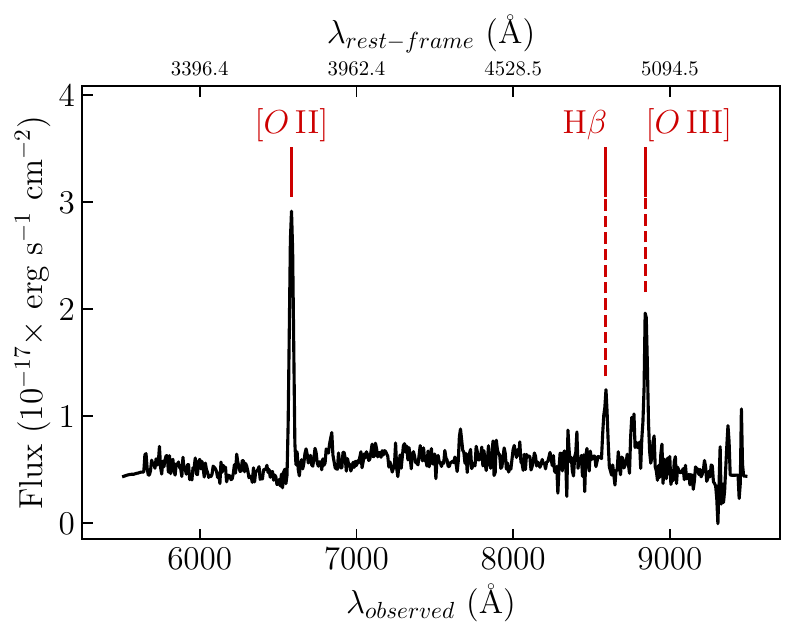}
    \caption{Spectrum of a source from our sample (VIPERS ID: 105161086; HELP ID: J021747.792-055145.509) at $z_{spec}=0.766$, highlighting the $\left[ \ion{O}{ii} \right] \lambda 3727$,  $\left[ \ion{O}{iii} \right] \lambda 4959, \lambda5007$, and $\text{H}\beta$ lines.
    }
    \label{fig:figure1}
\end{figure}

We used the penalized pixel fitting code (pPXF) \citep{ppxf} to model the VIPERS spectra by fitting stellar and gas templates from the MILES library \citep{Vazdekis2010}, as described in Pistis et al. (in prep.).  A single Gaussian was used to fit each emission line of the gas component, yielding integrated fluxes and errors. Table \ref{tab:table1} shows the median signal-to-noise ratios (S:N) of the resulting four emission lines of our sample of galaxies. An example of the spectrum of a galaxy from our sample is shown in Fig. \ref{fig:figure1}. Equivalent widths (EWs) and their errors were also computed.

\begin{table}
    \centering
        \caption{Signal-to-noise ratios of detected lines of our sample.}

    \begin{tabular}{c c}
    \toprule \midrule\rowcolor{lightgray!10}

     \ \ \ \ \ \ \  \ \ \ \  \ \ \ \ \   Line\ \ \ \ \ \ \  \ \ \ \  \ \ \ \  \   &  \ \ \ \ \ \ \  \ \   \ \ \ \ \ \ \  Median S:N\ \ \ \ \ \ \   \ \  \ \ \ \ \ \ \  \\
        \midrule\midrule
        $\left[ \ion{O}{ii} \right] \lambda 3727$ & 21.79\\
        $\text{H}\beta$ & 9.26\\
        $\left[ \ion{O}{iii} \right] \lambda 4959$ & 3.38 \\
        $\left[ \ion{O}{iii} \right] \lambda5007$ & 8.12\\
        \bottomrule
    \end{tabular}
    \label{tab:table1}
\end{table}

\subsection{Photometric data}\label{photo}

The fields surveyed by VIPERS were previously mapped by the Canada-France-Hawaii Telescope Legacy Survey (CFHTLS), in the optical range, with the Wide photometric catalog T0007 release \citep{Hudelot2012}. The T0007 catalog with $ugriz$ detections is an improvement of the initial T0005 and T0006 catalogs \citep{Guzzo2014}. In this newer catalog, the apparent magnitudes are given in the AB system, and were corrected for Galactic extinction, with an extinction factor derived at the position of each galaxy using the dust maps of \citet{Schlegel1998}. Near infrared magnitudes come from the $Ks$ band detection of CFHT/WIRCam.

We required our sample to have detections in the FUV and NUV bands of the GALEX deep imaging survey. In the catalog of \citet{Moutard2016}, GALEX data towards the CFHTLS-VIPERS fields were gathered, to measure the physical properties (e.g., SFR, stellar masses) of all galaxies in the VIPERS spectroscopic survey. The CFHTLS-T0007 maps were used as a reference to measure the Ks-band photometry, and the $u$ band-selected sources were used as priors in estimating the redshift, as well as the FUV and NUV photometry.\\

Additionally, we extended the wavelength coverage to the FIR by cross matching with the \emph{Herschel} Extragalactic Legacy Project \citep[HELP,][]{Shirley2021} catalog, which provides a unique statistical data of IR detection of millions of galaxies detected at long wavelengths with \emph{Herschel}, constructed homogeneously. To achieve the HELP catalog, \emph{Herschel} fluxes were estimated using XID+ pipeline \citep{Hurley2017}, a probabilistic deblender of SPIRE maps which takes into account the positions of sources detected with \emph{Spitzer} at 24 $\mu\text{m}$ \citep{Duncan2018} from the Multiband Image Photometer \citep[MIPS,][]{Rieke2004}. Detections from IRAC at its four channels from \emph{Spitzer} were added to our data. We used data from the PACS instrument at 100 and 160 $\mu\text{m}$, and from SPIRE at 250, 350, and 500 $\mu\text{m}$.
 
\subsection{Final sample}\label{final_sample}

Apart from the redshift flag selection, as discussed in Section~\ref{spectro}, we also restrain our sample to star-forming galaxies using the modified Baldwin, Phillips $\&$ Terlevich diagrams \citep[BPT,][]{BPT81} by \citet{lamareille2010spectral}. This modified BPT diagram allows us to use the lines available in VIPERS, to discard active galactic nuclei (AGNs), low ionization nuclear emission regions (LINERs), and Seyfert sources.

The above-described selection yields a sample of 1\,049 galaxies, with their photometric S:N described in Table \ref{tab:Table2}, covering a redshift range of $0.5 < z < 0.8$. A total of 592 galaxies in our sample have FIR detections from the HELP catalog. All of the galaxies in our sample have spectroscopic redshifts. Table \ref{tab:Table2} shows the photometric bands used for our data and the associated S:N for our final sample.
\begin{table*}
 \caption{Summary of photometric data in each band with its centered wavelength, the mean of S:N, and the number of detections in our sample.}
  \begin{center}
  \begin{tabular}{l c c c c c}
     \toprule\midrule\rowcolor{lightgray!10}

      Telescope/\ \ \ \ \ \ \ \ \ \ \ \ \ & \ \ \ \ \ \ \ \ \ Band \ \ \  \ \ \ \ \ \ & \ \ \ \ \ \ \ \ \ \ \ $\lambda$  \ \ \  \ \ \ \ \ \ \ \ & \ \ \ \ \ \ \ \ \ \ \ Median \ \ \ \  \ \ \ \ \ \ \ &   \ \ \ \ \ \ \ \ \ N\textsuperscript{\underline{o}} of\ \ \ \ \  \ \  \\ \rowcolor{lightgray!10}

 Instrument    &      & $(\mu m)$  &  S:N  &  detections\\
    \midrule\midrule
      GALEX             & FUV   & 0.15 & 5.45 & 1049\\
                        & NUV   & 0.23 & 9.78 & 1049\\
      \midrule
       CFHT/             & $u$   & 0.38 & 31 &  1049\\
       MegaCam           & $g$   & 0.49 & 61.69 &  1049\\
                         & $r$   & 0.63 & 50.62 &  1049\\
                         & $i$   & 0.76 & 78.69 &  1049\\
                         & $z$   & 0.89 & 40.90 &  1049\\
       CFHT/WIRCam       & $Ks$  & 2.14 & 25.25 &  1049\\                  
    \midrule
     Spitzer/            & ch1 & 3.56 & 27.66 &  710\\
     IRAC                & ch2 & 4.50 & 19.56 &  471\\
                         & ch3 & 5.74 & 10.43   &  112\\
                         & ch4 & 7.93 & 13.25   &  112\\
    \midrule
     Spitzer/MIPS        & MIPS1 & 24 & 15.90 &  116\\
                        &   MIPS2  &70&   10.23     & 18\\
                         & MIPS3   &160 &  6.19     & 3\\
    \midrule
     Herschel/           & 100~$\mu$m & 102.62 & 1.30 &  592\\
     PACS                & 160~$\mu$m & 167.14 & 1.38 &  592\\
    \midrule
     Herschel/           & 250~$\mu$m & 251.50 & 2.63 &  592\\
     SPIRE               & 350~$\mu$m & 352.83 & 1.51  &  592\\
                         & 500~$\mu$m & 511.60 & 1.05 &  592\\
    \bottomrule

     \end{tabular}
     \end{center}
     \label{tab:Table2}
\end{table*}

\begin{figure}
    \centering
    \includegraphics[width=\columnwidth]{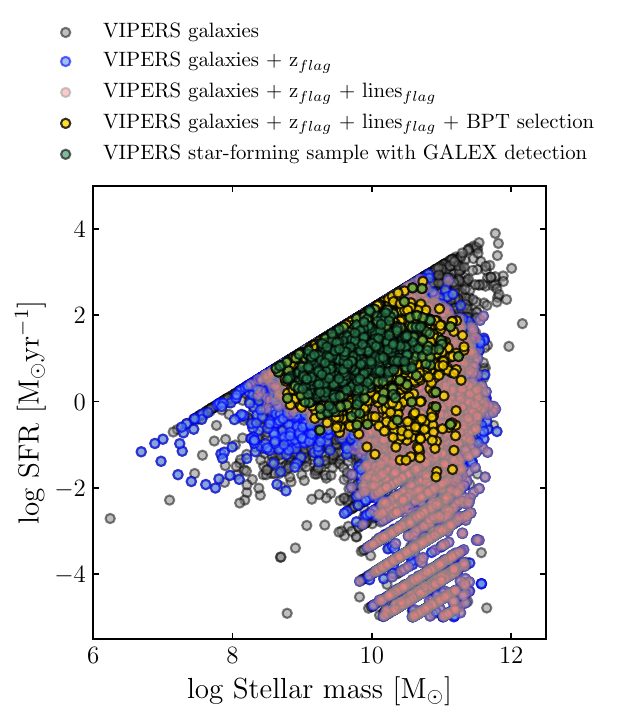}
    \caption{Distribution of galaxies considered in this work throughout the selection process in the main sequence plane of star-forming galaxies. The SFRs and stellar masses in this plot were computed for the initial VIPERS catalog using \texttt{HYPERZ} \citep[][]{Bolzonella2000}.
    }
    \label{fig:figure_cuts}
\end{figure}

Selection cuts were necessary to perform the analysis in this work, however, it results in few biases to take into account. Here we present the possible biases as well as their possible implications in our conclusions. The first selection to address is the selection of galaxies characterized by the strong emission lines, as those in majority, have the most reliable ($\geq99\%$ accuracy, $3< z_{flag}< 4.5$) redshift measurements in the VIPERS sample. Extremely dust obscured galaxies are removed with this cut, since their emission lines might be dim, as described in \citet{Scodeggio2018}.\\
Furthermore, the selection on the quality of the lines was needed to select the galaxies that are considered to have reliable line measurement in VIPERS \citep[e.g.,][]{Pistis2022, Figueira2022}. Also, to compute metallicity using the R$_{23}$ ratio, all three lines should be observed inside the spectral coverage of the instrument. For this reason we limited the sample to redshift $z_{max}\sim0.8$. For the quality of lines of our galaxies (Pistis et al. in prep.; Garilli, private communication), the line flags were taking into account the following criteria: the difference between the observed and Gaussian peak, the FWHM of the line, the difference between the observed and Gaussian amplitude, and the S/N. We adopted the minimum requirement for our sample for the first three criteria \citep[as detailed in][Pistis et al. in prep.]{Scodeggio2018, Pistis2022, Figueira2022} However, we did not limit ourselves to high S/N criteria on emission lines, but the redshift flag resulted in high S/N (presented in Table \ref{tab:table1}).\\
A key aspect of this work is to model in the most reliable way possible the short wavelength fluxes of our sample, in order to derive the $\beta$ values, as well as the attenuation values. We therefore required to have FUV and NUV detections for our sample to not over-fit this slope. Additionally, for almost half of our sample there was no available IR data, therefore we needed to rely on the energy balance of the SED models in order to estimate the IR luminosity of these objects, which required the highest coverage possible in the short wavelength to adopt, in the most correct way, the attenuation curve.  This selection cut removes naturally the dusty and IR-bright sources. 
The selection on the BPT diagram was needed in order to reliably measure the IRX-$\beta$ relation for star-forming galaxies. We therefore do not consider passive galaxies in this work. The distribution of our sample in the main sequence of star-forming galaxies after every selection cut is presented in Fig. \ref{fig:figure_cuts}. In Fig. \ref{fig:figure_cuts}, the SFR and stellar mass values were estimated in the initial catalog using \texttt{HYPERZ} \citep{Bolzonella2000, Davidzon2013}. These cuts excluded the low mass galaxies, as well as passive ones. These selections and the resulting sample sizes after every cut are shown in Table \ref{tab:table_selection}.

\begin{table}
    \centering
        \caption{Steps of data selection and the size of our sample after each selection step.}

    \begin{tabular}{l r}
    \toprule \midrule\rowcolor{lightgray!10}

      \  \ \ \ \  \ \ \ \ \   Selection\ \ \ \ \ \ \  \ \ \ \  \ \ \ \  \   &  \ \ \ \ \ \ \  \ \   \ \ \ \ \ \ \  Sample size\ \ \ \ \ \ \   \ \  \    \\
        \midrule\midrule
        VIPERS sample & 88\,340\\
        redshift flag & 53\,680\\
        lines flag &  36\,850\\
        BPT selection & 6\,251 \\
        GALEX detection & 1\,049\\
        \bottomrule
    \end{tabular}
    \label{tab:table_selection}
\end{table}

\section{Estimating physical properties from spectral energy distribution}\label{sed}
To model the spectra of our galaxies, we first assumed a stellar population. We then attenuated this stellar population with dust, assuming different dust and stars distribution. Also, we included the line measurements from the $\left[ \ion{O}{ii} \right] \lambda 3727$,  $\left[ \ion{O}{iii} \right] \lambda 4959, \lambda5007$, and $\text{H}\beta$ lines in the SED fitting procedure in \texttt{CIGALE} as done in \citet{Villa2021}. In the following subsections, we describe the different aspects of our SED fitting strategy, and the motivation of our choice of certain templates and parameters.

\subsection{Stellar population and its formation history}
To model the stellar emission and the spectral evolution of galaxies, we considered stars with different ages and a certain metallicity. In this work, we use the stellar population library of \cite{BC03}, a solar metallicity and an IMF of \cite{Chabrier2003}, which takes into account a single star IMF as well as a binary star systems. In terms of modeling, star formation histories (SFH) are very sensitive to many complex factors including galaxy environments, merging, gas accretion and its depletion \citep[e.g.][]{Elbaz2011,Ciesla2018,Schreiber2018,Pearson2019}. The SFHs have a significant effect on fitting the UV part of the SED, and consequently affecting the derived physical parameters such as the stellar masses and the SFRs. 
Initially, we tested different SFH models to fit the observed short wavelength photometry of our sources. These include a simple exponentially decreasing star formation characterized by an e-folding time $\tau$, and a similar delayed model with a recent exponential burst or quench episode \citep{Ciesla2017}. In our models, we gave the e-folding time of the main stellar population ($\tau_{main}$) a wide range of variation. Our initial results did not favor a recent burst in our sample, therefore we proceeded in using the simpler delayed SFH. In such star formation scenario, a galaxy has built the majority of its stellar population in its earlier evolutionary phase, then the star formation activity slowly decreases over time. The SFR evolution over time is hereby modeled with:
\begin{equation}
    \mathrm{SFR}(t)\ \propto\ \frac{t}{\tau^2}\ e^{-t/\tau}\,
\end{equation}
    where it translates into a delayed SFH slowed by the factor of $\tau^2$, the e-folding time of the main stellar population, and extended over the large part of the age of the galaxy. This SFH provided better fits compared to the one with the recent burst. We vary $\tau$ as shown in Table \ref{tab:Table3}, to give a comprehensive flexibility of the delayed formation of the main stellar population.

\begin{table*}
    \caption{Input parameters of the important physical models used to fit the SEDs of our sample with \texttt{CIGALE}.}
   \label{tab:Table3}
\centering
    \begin{tabular}{l  r}
\toprule\midrule 
    Parameter\ \ \ \ \ \ \  \ \ \ \ \ \ \  \ \ \ \ \ \ \ \ \ \ \ \ \ \  \ \ \ \ \ \ \  \ \ \ \ \ \ \ \ \ \ \ \ \ \ \ \ \ \ \ \ \  \ \ \ \ \ \ \  \ \ \ \ \ \ \ \ \ \ \ \ \ \  \ \ \ \ \ \ \ &  \ \ \ \ \ \ \ \ \ \ \ \ \ \  \ \ \ \ \ \ \  \ \ \ \ \ \ \ \ \ \ \ \ \ \  \ \ \ \ \ \ \  \ \ \ \ \ \ \ \ \ \ \ \ \ \  \ \ \ \ \ \ \ \ \ \ \ \ \ \  \ \ \ \ \ \ \  \ \ \ \ \ \ \ Values \\
\midrule\midrule
\rowcolor{lightgray!10}
     \multicolumn{2}{c}{ \ \ \ \ \ \ \  Star formation history \ \ \ \ \ \ \ }   \\
     \midrule     
\multicolumn{2}{c}{delayed}   \\
\midrule
     Stellar age [Gyr] & equally-spaced 32 values in [0.5, 8] \\
     e-folding time ($\tau$) [Gyr] & equally-spaced 22 values in [0.5, 6]  \\
\midrule\midrule
\rowcolor{lightgray!10}
    \multicolumn{2}{c}{Dust attenuation laws\tablefootmark{(a)}}   \\
\midrule
    \multicolumn{2}{c}{\citep{Calzetti2000}}   \\
\midrule
    Colour excess of young stars  E(B-V) mag & 10 values in [0.1, 1] \\
    Color excess of old stars (f$_{att}$)  & 0.1, 0.3, 0.5, 0.8, 1.0\\
\midrule
    \multicolumn{2}{c}{\citep{CharlotFall2000}}    \\
\midrule
    V-band attenuation in the ISM  (A$_V^{ISM}$) mag & 30 values in [0.3 - 3] \\
    Av$_{V}^{ISM}$ / (A$_V^{BC}+A_V^{ISM})$ & 0.1, 0.3, 0.5, 0.8, 1 \\
    Power law slope of the ISM &-0.7\\
    Power law slope of the BC &  -0.7 \\
\midrule\midrule
\rowcolor{lightgray!10}
\multicolumn{2}{c}{Dust emission}   \\
\midrule
    \multicolumn{2}{c}{\citep{Draine2014}}   \\
    \midrule
    Mass fraction of PAH &  1.12, 2.50, 3.90, 5.26,  6.63 \\
    Minimum radiation field (U$_{min}$) & 1, 5, 15, 25, 35 \\
    Power law slope ($\alpha$) & 2 \\
    Dust fraction in PDRs ($\gamma$) & 5 values in [0.01, 0.2]\\
    \midrule
\bottomrule
            \end{tabular}
        \tablefoottext{a}{One attenuation law was used at a time in the SED fitting procedures. Both approaches were used in order to derive accurate physical properties.}
       \end{table*}
\subsection{Dust attenuation}
Modeling dust attenuation is an indispensable priority in any panchromatic SED fitting in order to reproduce the short wavelength photometry and therefore to extract accurate physical properties.\smallbreak
In this work, we used two contrasting approaches of attenuation laws for our SED fitting: the approach of \citet[][]{Calzetti2000} and that of \citet[][]{CharlotFall2000}. While these two attenuation approaches are relatively simple, they differ on how to attenuate a given stellar population. Recently, \citet[][]{buat2019, Hamed21} showed that for statistical samples, it is necessary to use these two attenuation laws in order to recover the physical properties of galaxies, and chose the best fit model among the two.

The attenuation curve of \citet{Calzetti2000} is a screen dust model. It was tuned to fit a sample of starbursts in the local Universe, which represent high redshift UV-bright galaxies. This curve attenuates a stellar population with a simple power-law:
\begin{equation}
k(\lambda) = \dfrac{A(\lambda)}{E(B-V)},
\end{equation}
where k($\lambda$) is the attenuation curve at a certain wavelength, A($\lambda$) is the extinction curve, and $E(B-V)$ is the color excess between the $B$ and $V$ bands.

Despite its simplicity, this attenuation curve is widely used in the literature \citep[e.g.][]{Burgarella2005,Buat12,Malek2014,Malek2017,Pearson2017,Elbaz2018,Buat2018,Ciesla2020}. However, it does not always succeed in reproducing the UV extinction of galaxies at higher redshifts \citep{Noll2009,LoFaro2017}.\smallbreak
Another approach is to consider the dust spatial distribution relative to the stellar population. This is the core of the attenuation curve of \citet[][]{CharlotFall2000}. In this approach, dust is considered to attenuate the dense and cooler stellar birth clouds (hereafter BCs) differently than ambient diffuse interstellar media (ISM). This configuration is expressed by two independent power-laws:
\begin{equation}
A(\lambda)_{\text{ISM}} \propto \left(\dfrac{\lambda}{\lambda_{V}}\right)^{\delta_{ISM}} \text{and}\ \ A(\lambda)_{\text{BC}} \propto \left(\dfrac{\lambda}{\lambda_{V}}\right)^{\delta_{BC}},
\end{equation}
where $\delta_{ISM}$ and $\delta_{BC}$ are the slopes of attenuation in the ISM and the BCs respectively. Young stars that are in the BCs will therefore be attenuated twice: by the surrounding dust and additionally by the dust in the diffuse ISM. CF00 found that $\delta_{ISM}$ = $\delta_{BC}$ = -0.7 satisfied dust attenuation in nearby galaxies, however, this curve is frequently used at higher redshifts \citep[e.g.][]{Malek2018,Buat2018,Salim2020,Donevski2020,Hamed21}. By attenuating at higher wavelengths (until the NIR) more efficiently than the recipe of \citet{Calzetti2000}, this approach considers a more attenuated older stellar population. This is due to the fact that the double power-law attenuation recipe of \citet{CharlotFall2000} attenuates the ISM and the BC separately, where the ISM is assumed to contain the older stars. This attenuation curve, as a consequence, is shallower in the optical-to-NIR wavelength range than that of \citet{Calzetti2000}, and this might produce slightly larger stellar masses \citep{Malek2018, buat2019, Hamed21, Hamed23}. To model dust attenuation of the galaxies of our sample, we use the aforementioned laws, with the parameters presented in Table \ref{tab:Table3}.

\subsection{Dust emission}
Modeling dust emission of our sample is not only crucial to derive the key observables that govern galaxy evolution such as the SFRs, but it is a cornerstone in correctly deriving IR luminosities which in turn might affect the estimation of the IRX. To reproduce the IR emission in our SED models, we use the templates of \citealt{Draine2014}. These templates take into consideration different sizes of grains of carbon and silicate, hence, allowing different temperatures of dust grains. They rely on observations and are widely used in the literature to fit FIR SEDs \citep[e.g.,][]{buat2019,Burgarella2020, Hamed21}.\\
\begin{figure}
    \centering
    \includegraphics[width=1\columnwidth]{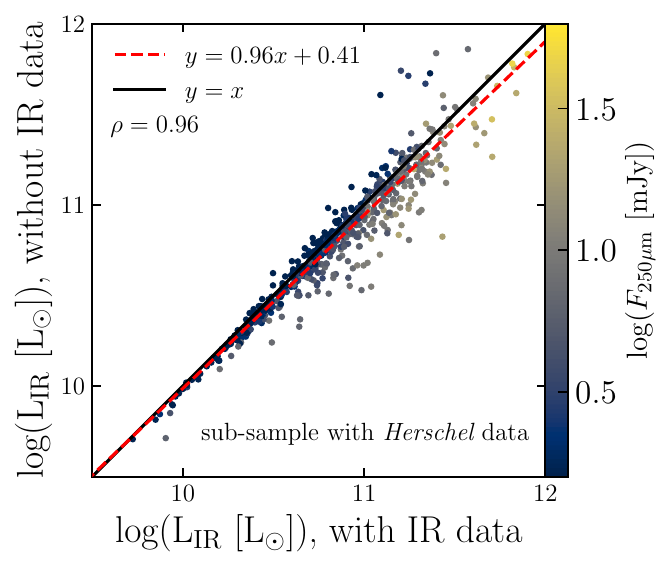}
    \caption{Dust luminosities estimated for the IR-detected sub-sample (592 galaxies out of 1049), with and without IR data. The solid black line shows the one-to-one relation. The dashed red line shows the best fit of this comparison, and $\rho$ is the Pearson coefficient. The scatter is color-coded with the fluxes of the sub-sample at 250$\mu$m.}
    \label{fig:dust_luminosity}
\end{figure}
Given that $\sim$56$\%$ of our sample possess FIR data from the HELP catalog, we fit these IR-detected galaxies with \citet{Draine2014} dust templates, with and without IR data. That is, in one fitting, we used the full available panchromatic photometric coverage from FUV to 500~$\mu$m, using a stellar population library, SFH, dust attenuation and dust emission templates. In another fitting procedure we fit with the same templates only the selected range of photometry from FUV to IRAC4 band (8~$\mu$m). We then compare dust luminosities estimated using these two techniques, this comparison is provided in Figure \ref{fig:dust_luminosity}. In this Figure, the sources are color-coded with the intensity of the flux at 250\,$\mu$m. We notice that IR luminosities estimated solely on the energy balance based on the UV-NIR detections, are slightly underestimated for IR brighter objects. Although the gradient of the change is not large ($0.2\, <\,\mathrm{log}(\mathrm{F}_{250\mu m}\,[\mathrm{mJy}])\,<\,1.7$\,), the IR luminosities derived from the energy balance for the galaxies that miss IR detection can be slightly overestimated. Similar results were observed for the same test in \citet{Malek2018} for star-forming galaxies with the redshift ranging from $0.5<z<2$, and for dusty star-forming galaxies at $z\sim~2$ in \citet{buat2019}, and in \citet{Junais23} for low and high surface brightness galaxies. In order to estimate the IR luminosities of the objects in Figure \ref{fig:dust_luminosity}, the best attenuation law that describes each source was used, based on the reduced $\chi^2$.\\
The obtained IR luminosities with the aforementioned techniques correlated with a Pearson coefficient of $\rho=0.96$. We then proceeded with fitting the photometry of galaxies that do not possess IR observations in the HELP catalog (457 galaxies out of 1049) with the same input parameters of \citet{Draine2014} templates, computing the IR luminosities based on the energy balance applied in the SED fitting using the short wavelength data. This allows for a reliable estimation of IR luminosities. We show the star-forming galaxy main sequence of our sample in Figure \ref{fig:MS}. Our fitting technique provided an overall low uncertainties of physical properties, such as the stellar masses and the SFR.\\
\begin{figure}
    \centering
    \includegraphics[width=1\columnwidth]{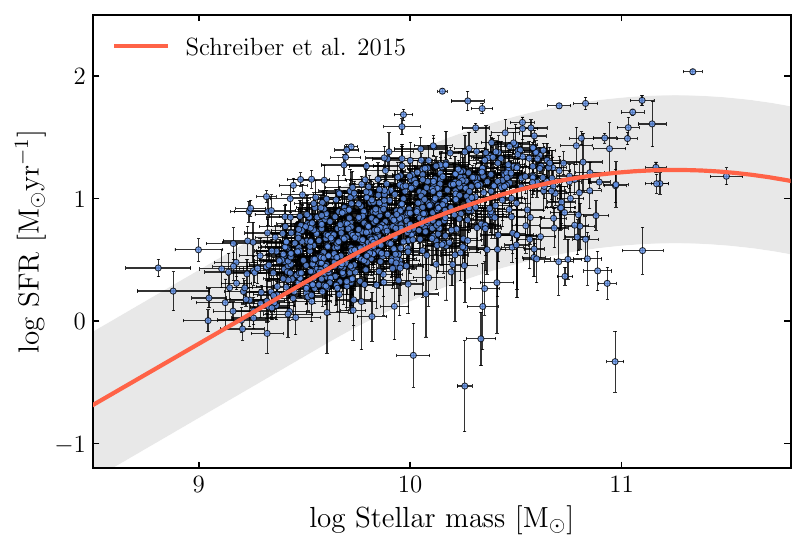}
    \caption{Star formation rates of our sample fitted with the best attenuation curve against stellar masses. The red line shows the main sequence of star-forming galaxies from \citet[][]{Schreiber15} at $z=0.6$ (the median of our sample).}
    \label{fig:MS}
\end{figure}

\subsection{SED quality, model assessment}\label{mock}
\begin{figure}[t!]
    \centering
    \minipage{0.25\textwidth}
    \includegraphics[width=\linewidth]{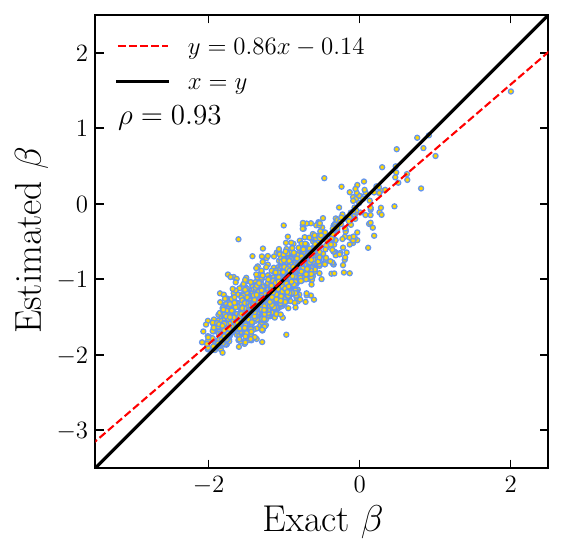}
    \endminipage
    \minipage{0.25\textwidth}
    \includegraphics[width=\linewidth]{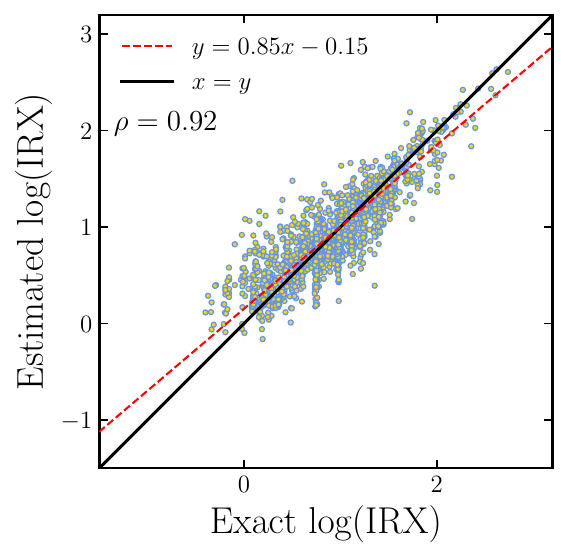}
    \endminipage\\
    \minipage{0.251\textwidth}
    \includegraphics[width=\linewidth]{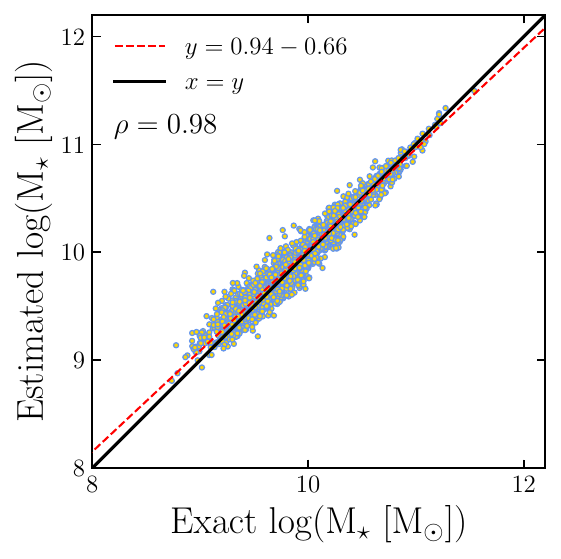}
    \endminipage
    \centering
    \minipage{0.249\textwidth}
    \includegraphics[width=\linewidth]{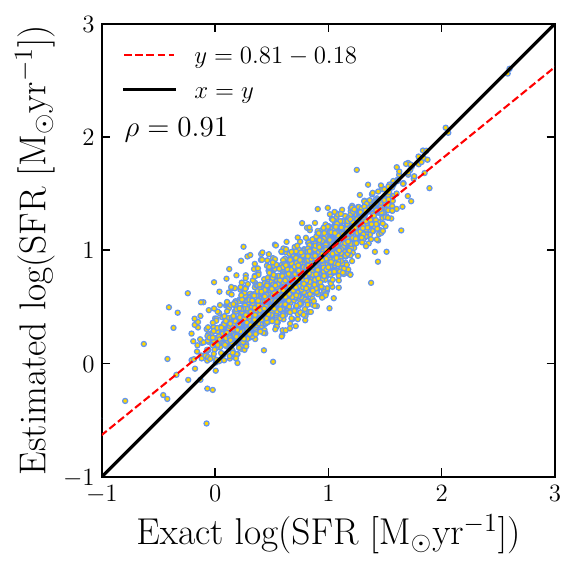}
    \endminipage
    \caption{\footnotesize Comparison between some of the true parameters of the mock SEDs and the results from the SED modeling with CIGALE. The "exact" values are from mock models. $\rho$ is the Pearson coefficient.}
       \label{fig:mock}
\end{figure}

To test the reliability of our SED models, with \texttt{CIGALE} we generated a mock galaxy sample and fitted SEDs with the same methods applied to our sample. We build the mock catalog by perturbing the fluxes of our best SEDs with errors sampled from a Gaussian distribution with a standard deviation equivalent to the uncertainty observed in the real fluxes. We show the mock analysis of the important estimated quantities in Figure \ref{fig:mock}, where we show the comparison between the real physical properties that we derived for our sample and its mock equivalent. The mean $\chi^2_{reduced}$ of our sample was 8.6, mainly due to the large error bars of the IR data. For each galaxy, we selected the attenuation law that best describes its observed photometry by comparing their $\chi^2$, as in \citet[][]{buat2019, Hamed21}.\\

To estimate $\beta$ for each galaxy in our sample, we fit a power-low function to the SED of each galaxy between the rest-frame ranges 0.126 $\mu$m and 0.260 $\mu$m. IRX was estimated directly from the SED fitting process, by dividing the IR luminosity by the FUV luminosity for each source, that is, IRX =log(L$_{IR}$/L$_{FUV}$).\\
We also compared the SFRs derived using the spectroscopic lines, notably the  H$\beta$ and $\ion{O}{ii}$ lines, with the ones obtained using SED of the photometry. We show this comparison in Appendix \ref{appendixa}. This ensures a coherent interpretation of the SFR using the different methods. Overall, the mock analysis showed that our estimations of the physical observables, such as dust luminosities, SFRs, stellar masses, and IRX and $\beta$, are reliable.

\section{Estimating the metallicity}\label{metal}
Different calibrations are often used in the literature to measure the gas phase metallicity based on different line ratios \citep{pilyugin2001, tremonti2004origin, nagao2006gas, curti2016new}. Making use of the robust oxygen emission lines along the H$\beta$ line, we measured the gas phase metallicity using the ratio
\begin{equation}
    R_{23} = \frac{\left[ \ion{O}{ii} \right] \lambda 3727 + \left[ \ion{O}{iii} \right]\lambda\lambda 4959, 5007}{\text{H}\beta}.
\end{equation}
The $R_{23}$ ratio was initially proposed in \citet[][]{Pagel1979}, and since then its tuning to oxygen abundance was improved by various photoionization models \citep[][]{nagao2006gas}. To derive the metallicity of our galaxies, we used the calibration proposed by \citet[][]{tremonti2004origin}, which is based on the $R_{23}$ ratio of star-forming galaxies of double-valued $R_{23}$-abundance ratios. Following \cite{nagao2006gas} we verified if our sample is in the upper branch ($\left[ \ion{O}{iii} \right]\lambda 5007 / \left[ \ion{O}{ii} \right] \lambda 3727 < 2$), and found that all data of our sample belong to the upper branch. Additionally, we removed all the sources that fall below the calibration limit of $12 + \nolinebreak \log \left( \text{O/H} \right) \leq \nolinebreak  8.4$, since this metallicity calibration is valid for values above that threshold. This discarded 47 galaxies, and left a total of 1002 galaxies for which valid metallicity estimation was calculated.\\
The calibration proposed by \citet[][]{tremonti2004origin} estimates the metallicity from theoretical model fitting of emission-lines. The model fits are calculated combining SSP synthesis models from \cite{BC03} and \texttt{CLOUDY} photoionization models \citep{Ferland98}. The relation between metallicity and $R_{23}$ is given by :
\begin{equation}
    12 + \log \left( \text{O/H} \right) = 9.185 - 0.313 x - 0.264 x^2 - 0.321 x^3,
\end{equation}
where $x \equiv \log R_{23}$.\\

To estimate the metallicities of our sample using the aforementioned lines, we corrected the lines for attenuation. The value of attenuation A$_{(V, stellar)}$ was computed via SED fitting. We then passed from A$_{(V, stellar)}$ to A$_{(V, nebular)}$ assuming factor $(f=0.57)$ as proposed in \citet{rodriguez2022}. To correct each line for attenuation we applied the law described in \citet{Cardelli89}.

\section{IRX-$\beta$ relation of our sample}\label{irx_beta}

\begin{figure}[t!]
    \centering
    \includegraphics[width=1\columnwidth]{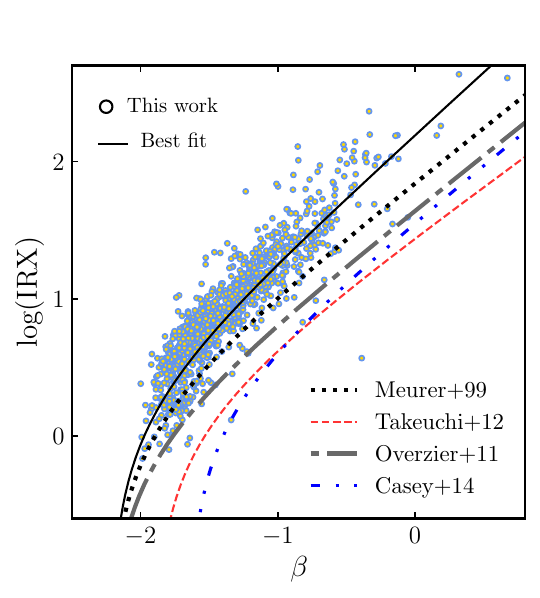}
    \caption{IRX-$\beta$ scatter of our sample. The original \citet[][]{Meurer99} fit is shown in dotted line. We also compare with fits from \citet{Overzier2011, Takeuchi2012, Casey2014}, denoted with grey dash-dotted, red dashed, and dash-double dotted lines, respectively. The solid black line shows Equation \ref{eq-fit}, describing our sample.}
    \label{fig:irx_beta}
\end{figure}

We show the scatter of IRX-$\beta$ of our sample of star-forming galaxies in Figure \ref{fig:irx_beta}. This scatter is observed to be higher than the relation fitted on the sample of local starburst galaxies by \citet{Meurer99}. We fit the scatter of our sample in the IRX-$\beta$ diagram following the same method presented in \citet{Hao2011} and used similarly in \citet{Boquien2012}.\\

To separate the influence of the SFH from the attenuation, we connect the attenuation to $\beta$:
\begin{equation}
    A_{\mathrm{FUV}} = (\beta - \beta_0) \times c_{\beta},
        \label{eq-beta}
\end{equation}
where $A_{\mathrm{FUV}}$ is the attenuation in the FUV band, $\beta_0$ is the intrinsic UV slope (without dust), and $c_{\beta}$ is the attenuation constant connecting both sides of the equation. The former can be seen as the degree to which the attenuation in the FUV band is influenced by the reddening caused by dust. On the other hand, IRX can be linked with dust attenuation via:
\begin{equation}
    A_{\mathrm{FUV}} = 2.5\times \mathrm{log} (1 + c_{ \mathrm{IRX}}10^{ \mathrm{IRX}}),
    \label{eq-irx}
\end{equation}
where $c_{IRX}$ is the proportion of emission in the FUV band relative to the attenuation observed in other bands \citep{Meurer99, Boquien2012}. Equations \ref{eq-beta} and \ref{eq-irx} can be integrated as in \citet{Hao2011}:

\begin{equation}
    \mathrm{IRX} = \mathrm{log}\left( \frac{10^{0.4\times (\beta-\beta_0)c_{\beta}}-1}{c_{ \mathrm{IRX}}}\right).
    \label{eq-irx-beta}
\end{equation}

which, for our sample, gives the following equation: \begin{equation}\mathrm{IRX} = \mathrm{log}[(10^{0.91\beta+2.02}-1)/0.67].
\label{eq-fit}
\end{equation}
Our fitted relation lies above the relation given originally by \citet{Meurer99}, and the subsequent work on the same sample by \citet{Takeuchi2012}. Our relation is 0.22 dex higher on average than that of \citet{Meurer99}, with a mean difference being 0.11 dex for $-2.2< \beta <-0.8$ and 0.33 dex for  $-0.8< \beta <-0.6$. This can be attributed to different factors, such as the larger sample size in our work (Meurer+99 sample was 44 galaxies, \citet{Takeuchi2012} was 57 starburst galaxies), to the larger dynamical range of properties that our sample has or, simply, different redshift range which probe different epochs of Universe. For instance, Meurer+99 had only four data points at $\beta>-0.5$, and all of them are in the local Universe. Although slightly higher, our sample represents the normal star-forming galaxies at intermediate redshift, unlike the aforementioned works which studied primarily local UV-bright starbursts. Additionally, our sample might have higher specific dust masses than that of local galaxies, which helps in shifting our sources towards bluer beta and slightly higher IRX. Our fitted relation falls in between the ones derived for the local starbursts on the one hand,  and the high redshift dusty star-forming galaxies on the other hand \citep[][2<z<3.5]{Alvarez2016},  \citep[][2<z<3]{McLure2018}, \citep[][z=0 and 2<z<3]{Safarzadeh17}. Additionally, the fact that we applied the best attenuation law that describes every galaxy in our sample, has a natural impact on the loci of these galaxies in the IRX-$\beta$ plane \citep{Boquien2012}. The IRX-$\beta$ relation was shown to be related to the geometry of dust and stars \citep{Boquien2009, Boquien2012, Narayanan2018, Liang21}. Therefore, we consider our fit to be representative of the normal star-forming galaxies at intermediate redshift.

\begin{figure}
    \centering
    \includegraphics[width=0.9\columnwidth]{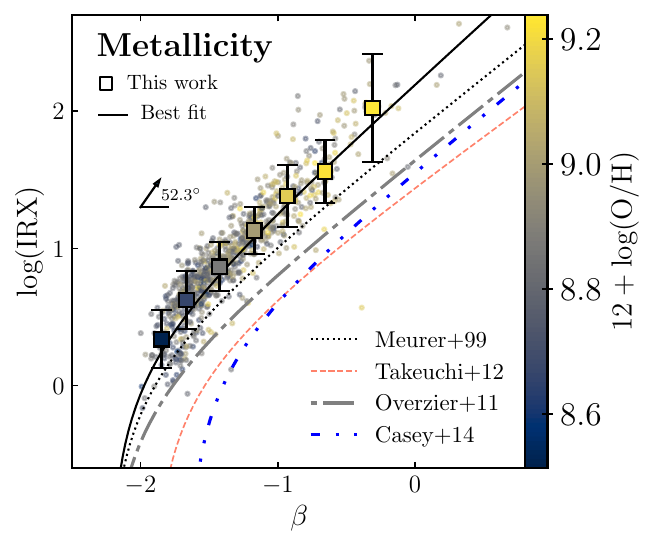}
    \caption{IRX-$\beta$ scatter of our sample color-coded with metallicity. Points denote the VIPERS sample binned by $\beta$, every bin contains the same number of galaxies (160 galaxies) except the last bin that contains 89 galaxies. The best fits from the literature and the best fit of our sample are also shown. Colors and line-styles of these fits are the same as in Fig. \ref{fig:irx_beta}. The arrow represents the trend of the metallicity, based on the scattered distribution (the direction is from the mean of the bottom 30$^{th}$ percentile of the distribution of metallicity to that of the top 70$^{th}$ percentile). The angle is between the arrow and the x axis.}
    \label{fig:irx-beta-metal}
\end{figure}

\section{Results \& Discussion: How do physical properties correlate with the IRX-$\beta$ relation?}\label{discussion}

\subsection{Correlation with metallicity}
A correlation between dust reddening and gas-phase metallicity is expected, as metals in the ISM will be depleted onto the same dust grains that attenuate the young stars and their emission lines. This means that the more metal content there is in the ISM, the more dust grains will be formed. Our findings suggest that metallicity is a significant contributing factor to the variability observed in the IRX-$\beta$ relation, as shown in Figure \ref{fig:irx-beta-metal}. Metallicity is indicative of the characteristics of small dust particles that affects the attenuation curve \citep{Zelko2020}. Lower gas-phase metallicity results in stronger UV radiation in the ISM, leading to the breaking down of larger dust grains into smaller particles, assuming a conservation of mass of dust. This results in an increase of the overall attenuation \citep{Shivaei2020}. Other works \citep[e.g.,][]{Conroy2010,Wild2011} have also found that higher gas-phase metallicity is indicative of higher dust attenuation.\\

Although higher dust-mass galaxies occupied the upper part of the scatter, the correlation was not net as with the metallicity. Therefore dust mass was not found to be a key factor in the diagram. However, we found that dust-to-star geometry was strongly responsible in moving the scatter to the right \citep{Popping2017}. Galaxies that preferred the attenuation curve of CF00 were located towards redder $\beta$ (higher $\beta$ values).

\subsection{Correlation with other galaxy properties}
\begin{figure*}
    \centering
    \includegraphics[width=1.9\columnwidth]{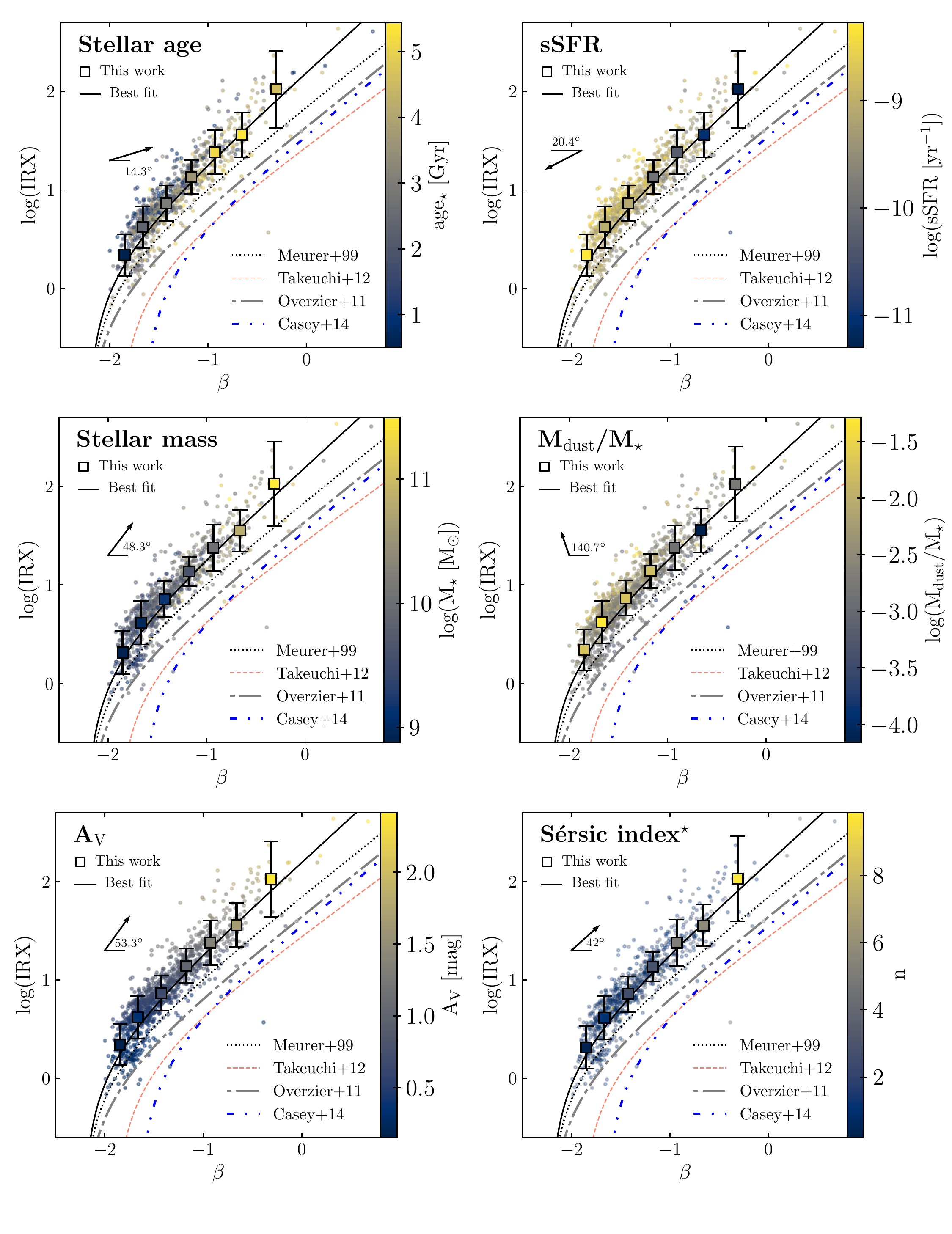}
    \caption{IRX-$\beta$ binned scatter of our sample color-coded with different physical properties. Squares denote the VIPERS sample binned by $\beta$, every bin contains the same number of galaxies (160 galaxies) except the last bin that contains 89 galaxies. The original \citet[][]{Meurer99} fit is shown in dotted line. We also compare with fits from \citet{Overzier2011, Takeuchi2012, Casey2014}, denoted with grey dash-dotted, red dashed, and dash-double dotted lines, respectively. The solid black line shows Equation \ref{eq-fit}, describing our sample. The arrows represent the direction of the trend, starting from the mean of the bottom 30$^{th}$ percentile of the distribution up to that of the top 70$^{th}$ percentile of a given quantity, they are meant to show the general direction of the trend. The lower right Figure with the S\'ersic index, a total of 823 galaxies (instead of 1049) are binned, due to \texttt{GALFIT} flag selection, based on \citet{Krywult2017}. The angles are between the arrows and the x or y axis.}
    \label{fig:mozaique}
\end{figure*}

We show the dependence of the IRX-$\beta$ relation with the physical properties of our sample in Figure \ref{fig:mozaique}. Higher mass galaxies with older stellar population occupy the higher IRX and $\beta$ values of the diagram. The Pearson coefficients between the stellar mass and IRX and $\beta$ are found to be $\rho=0.96$ and $\rho=0.98$ respectively for our sample, signifying a strong correlation. This almost linear scaling between the stellar mass and IRX and $\beta$ was also found at different redshift ranges \citep[e.g.,][]{Koprowski2018, Shivaei2020}.\\

On the other hand, the specific SFR (sSFR = log SFR/M$_{\star}$) is found to decrease at higher $\beta$. This decrease is found to be stronger with $\beta$ rather than IRX. The sSFR can be seen as a competition between the SFR and the stellar mass. At higher $\beta$ values, the sharper increase of the stellar mass overcomes the slow increase in the SFR of our sample. Generally, high SFR in galaxies, with higher dust masses, results in a larger fraction of young massive stars in the stellar population, which increases the UV radiation. This UV radiation is absorbed by dust, which re-emits in in the IR causing a higher IRX value.\\
\begin{figure}[b!]
    \centering
    \includegraphics[width=1\columnwidth]{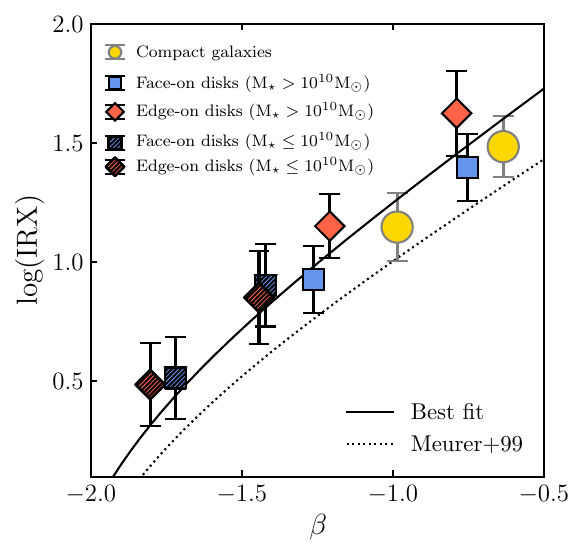}
    \caption{Our sample binned by $\beta$. Yellow circles represent compact galaxies ($n>2$). Blue squares show face-on disks ($b/a>0.5$ and $n<2$). Red diamonds represent edge-on disk galaxies ($b/a<0.5$ and $n<2$). Full markers show the higher mass galaxies in our sample, that is M$_{\star}$>10$^{10}$[M$_{\odot}$]. While the hatched markers are that of the lower mass sources, that is M$_{\star}$ $\leq$ 10$^{10}$[M$_{\odot}$}].
    \label{fig:b/a}
\end{figure}
The attenuation in the $V$ band shows a strong trend in the diagram, with more optically attenuated galaxies located in the higher IRX and high $\beta$ values. The specific dust mass (M$_{\mathrm{dust}}$/M$_{\star}$) correlation with IRX-$\beta$ is shown in Figure \ref{fig:mozaique} (middle right panel). With higher specific dust mass, galaxies move away from the fitted relation towards lower $\beta$ values.\\
\subsection{Variation with morphology}
The morphological parameters of our sample were estimated using \texttt{GALFIT} in \citet{Krywult2017} where they applied S\'ersic profiles fitting over the $r$-band of CFHT in their sample of VIPERS galaxies. A full description of the methodology used in deriving morphological parameters is detailed in \citet{Krywult2017}.  We notice a strong correlation between the S\'ersic index ($n$) and the distribution of our sample in the IRX-$\beta$ diagram. the S\'ersic index gives a Pearson coefficient of 0.95 with both IRX and $\beta$ separately. We show the correlation with the S\'ersic index in Figure \ref{fig:mozaique} (lower right panel). The number of sources in this particular plot is less than the other plots of IRX-$\beta$, because we applied the most secure flag of S\'ersic profiles fitting. This flag discards the galaxies for which \texttt{GALFIT} did not converge, and keep galaxies that had $0.2<n<10$. This discarded 226 sources from our sample, and left 823 galaxies out of 1049. Our results show a strong correlation of the loci of our galaxies in the IRX-$\beta$ diagram with galaxy compactness in the optical band.\\

\begin{figure}[t!]
    \centering
    \minipage{0.5\textwidth}
    \includegraphics[width=\linewidth]{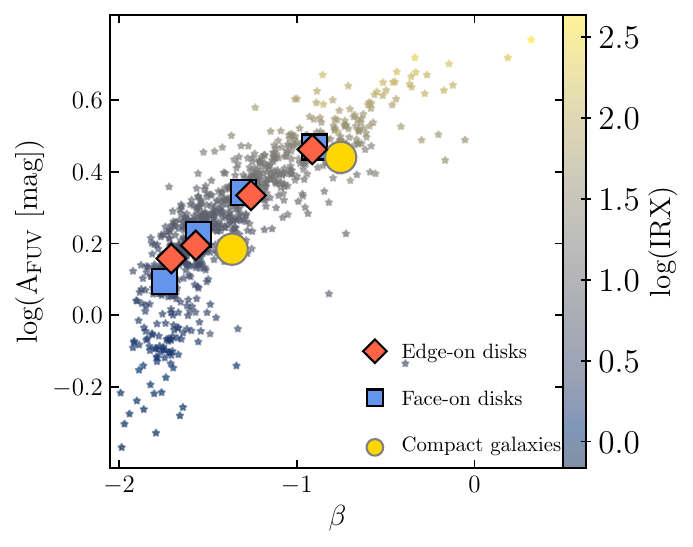}
    \endminipage
  
    \caption{\footnotesize Variation of the attenuation in FUV with $\beta$, color-coded with log(IRX). Edge-on disks have $b/a < 0.5$ and $n < 2$, while face-on disks have $b/a > 0.5$ and $n < 2$. Compact galaxies are those for which $n>2$.}
       \label{fig:b2a_properties}
\end{figure}    

Equivalently, we studied the variation of the loci of galaxies in the IRX-$\beta$ scatter with galaxy inclination as in \citet{Wang2018}. We used the same definition of disk galaxy inclinations as in the literature: galaxies for which the ratio of minor to major axis ($b/a$) > 0.5 are classified as face-on galaxies. On the other hand, galaxies whose $b/a \leq 0.5$ are considered edge-on galaxies. These definitions are valid for disk sources ($n < 2$). In our analysis, we also adopt the same notion of compact galaxies, as those for which $n>2$ \citep[][]{Wang2018}. We show the IRX-$\beta$ distribution of our sample based on galaxy morphology in Figure \ref{fig:b/a}, where every bin contains the same number of galaxies ($\sim110$ galaxies). Additionally, we separated the galaxies in our sample according to their stellar masses.\\

We find that compact galaxies ($n>2$) occupy the higher IRX values relative to the less compact ones. This is shown using the whole sample in Figure \ref{fig:mozaique} (lower right panel). However, for the disk sub-sample, we find that for galaxies whose stellar masses are M$_{\star}$>10$^{10}$[M$_{\odot}$] (259 sources), the IRX values are higher for edge-on galaxies than that of face-on galaxies. On the other hand, the lower-mass galaxies (M$_{\star}$ $\leq$ 10$^{10}$[M$_{\odot}$]) show no preferential loci on the IRX-$\beta$ diagram with inclination, as shown in Figure \ref{fig:b/a}, where similar results were shown in \citet{Wang2018}. The higher IRX values for high-mass edge-on galaxies, might be caused by the increase of the optical depth in these sources compared to the face-on galaxies \citep{Wang2018}. \\

In Figure \ref{fig:b2a_properties} we show the variation of the attenuation in FUV with $\beta$ for the different morphological classifications. For a fixed $\beta$ value, we find no difference in attenuation between edge-on and face-on galaxies. This suggests that the inclination of galaxies does not affect the IRX-$\beta$ scatter. Similar conclusion was also found in \citet{Wang2018}.

\section{Dust attenuation and galaxy environment}\label{env}
The density field of VIPERS data was computed by \citet{Cucciati2017}, where they estimated the local environment around the galaxies. In their work, they computed the galaxy density contrast between the local density at a comoving distance around each source, and the mean density at a given redshift. The mean density was achieved using mock catalogs. The local density of each galaxy was measured based on its fifth nearest neighbour.\\
The local density contrast is defined as \citep[e.g.,][]{Siudek2022}:
\begin{equation}
    \delta(RA, DEC, z) = \frac{\rho(RA, DEC, z)-\overline{\rho(z)}}{\overline{\rho(z)}},
    \label{eq_overdensity}
\end{equation}
where  $\rho(RA, DEC, z)$ is the local density of a given galaxy, and $\overline{\rho(z)}$ is the mean density at a redshift $z$.\\
\begin{figure*}
    \centering
    \minipage{\textwidth}
    \includegraphics[width=0.488\columnwidth]{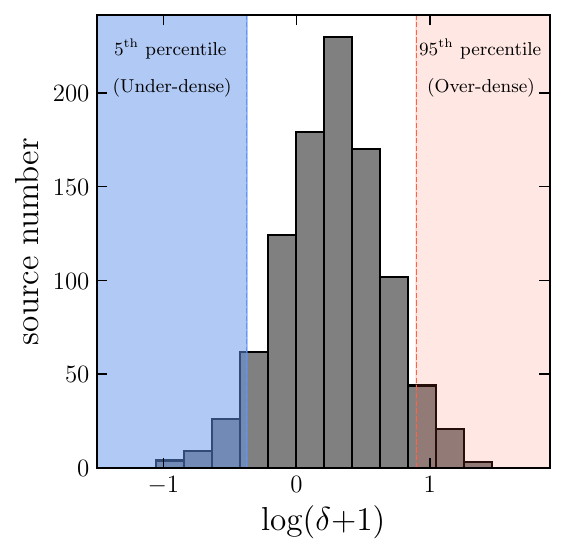}
    \includegraphics[width=0.512\columnwidth]{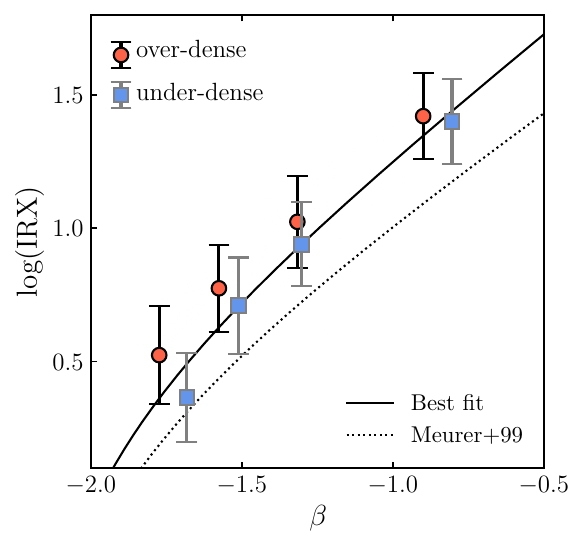}
    \endminipage
    \caption{\emph{Left panel:} Distribution of galaxy overdensity log($\delta$+1). In blue region we show the selection of the 5$^{th}$ percentile of the overdensity (the galaxies residing in less-dense environments), and in the red region the selection of the 95$^{th}$ percentile of the distribution (galaxies residing in overdense regions). \emph{Right panel:} The 5$^{th}$ percentile and the 95$^{th}$ percentile of the selection from the left panel of this figure, in the IRX-$\beta$ scatter.}
    \label{fig:overdensity}
\end{figure*}

We separate galaxies that reside in overdense and under-dense regions, defined as the 5$^{th}$ and 95$^{th}$ percentiles of the log(1+$\delta$) distribution, respectively (shown in Fig.~\ref{fig:overdensity}, left panel). We do this in order to see the subtle difference in the IRX-$\beta$ diagram (Figure \ref{fig:overdensity} right panel). We find no significant correlation with the IRX-$\beta$ relation. Even though the galaxies that reside in less dense environments are slightly shifted towards the higher $\beta$ values, for a given $\beta$, IRX is similar in the two galaxy groups within the error bars.\\

\begin{figure}
    \centering
    \includegraphics[width=1\columnwidth]{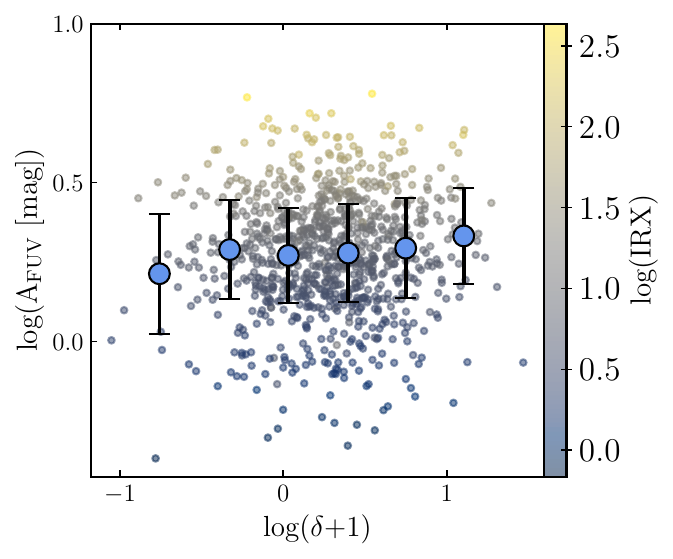}
    \caption{Attenuation in the FUV band, as a function of galaxy overdensities. The scatter represents the galaxies color-coded with IRX. We show the binned values for the whole sample (170 galaxies per bin).}
    \label{fig:env_fuv}
\end{figure}

We analyzed the attenuation in FUV bands of the galaxies from our sample with their environment overdensities. We show this in Figure \ref{fig:env_fuv}. We find no relation between the environment overdensities and dust attenuation. Galaxies' environments strongly affect their physical properties \citep[e.g.,][]{Peng2010}, however it is not understood how it might affect dust attenuation in galaxies. \citet{Shivaei2020} concluded similarly that environment does not seem to correlate with dust attenuation and consequently with the IRX-$\beta$ relation. We extend this conclusion from the high redshift in their work down to intermediate redshift, showing that the variation of the A$_{FUV}$ with galaxy overdensities is absent.\\

To quantify the degree of correlation in the IRX-$\beta$ diagram, we show in Figure \ref{fig:mozaique} the angles of the distribution of each physical property. We find that the most correlated quantities with the IRX-$\beta$ plane, are the metallicities, galaxy compactness, and stellar mass.\\

\section{Conclusions}\label{conclusions}
In this work, we dissected the IRX-$\beta$ dust attenuation relation for a large sample of 1049 galaxies at intermediate redshift ($0.5<z<0.8$). Having robust emission lines measurements, specifically the H$\beta$, $\ion{O}{ii}$ and the double [O{\,\sc{iii}}] lines, we estimated the gas-phase metallicities of 1002 sources in our sample using \citet{tremonti2004origin} calibration. Additionally, having full FUV to FIR detections of $\sim$ half of our sample, we computed reliably the SEDs of the galaxies. We showed that for galaxies that do not possess FIR detections, we can reliably estimate the dust luminosity based on the SED energy balance principle.\\

Among the tested physical properties of our sample in the IRX-$\beta$ relation, gas-phase metallicity correlated strongly with the loci of the galaxies within this diagram by moving the galaxies along the track of the relation, as was shown in Figure \ref{fig:irx-beta-metal}. A strong trend was also found with the stellar masses of our galaxies, and the age of their main stellar population. We conclude that metallicity strongly correlates with the  IRX-$\beta$ scatter, this also results from the older stars and higher masses at higher beta values. Galaxies with higher metallicities show higher IRX and higher beta values. The sSFR was found to decrease at lower $\beta$ values, due to the competition between SFR and stellar mass where the SFR were not high enough to compensate the clear increase of the stellar mass at higher IRX values. Similar results were achieved in the literature \citep[e.g.,][]{Boquien2012, Koprowski2018, Shivaei2020}. Our results suggest that at intermediate redshift, where earlier star-forming galaxies evolve through, the IRX-$\beta$ scatter remains dependent on these physical properties.\\

We notice that the correlation with the specific dust mass is strong in shifting the galaxies farther away from the IRX-$\beta$ relation towards lower $\beta$ values. This is due to the strong increase in the relative mass of dust compared to the stellar mass, leading to higher IR luminosities.\\

Having the morphological parameters of our sample \citep{Krywult2017}, we analyzed the effect that certain morphological properties have on the location of galaxies in the IRX-$\beta$ plane. We find a strong correlation with the S\'ersic index $n$, that is, with higher IRX and $\beta$ values, galaxy compactness seems to increase with the same rate. Morphologically, we find that more optically-compact objects (compact stellar population regions), witness a larger amount of attenuation that less compact galaxies.\\

We also tested the effect of galaxy inclination in the observed IRX-$\beta$ scatter, by separating our sample in edge-on disks, face-on disks and compact galaxies. Additionally, we split the sample into lower mass galaxies (M$_{\star}$ $\leq$ 10$^{10}$[M$_{\odot}$]) and higher-mass galaxies (M$_{\star}$>10$^{10}$[M$_{\odot}$]). Similarly as in \citet{Wang2018}, we find that for the higher mass sub-sample, IRX values were higher for edge-on disks. However, for the lower mass sub-sample, this variation is not noticeable. The increase in IRX values for edge-on higher mass galaxies could be attributed to the higher optical depth in these galaxies as compared to the face-on galaxies, as suggested by \citet{Wang2018}. Checking the variation of the attenuation in the FUV band with $\beta$, we found no difference in the amount of attenuation and $\beta$ with IRX, as was shown in Figure \ref{fig:b2a_properties}.\\

We studied the effect of galaxy environments on dust attenuation in Section \ref{env}. To do so, we took the 5$^{th}$ percentile and the 95$^{th}$ percentile of the distribution of our sample's overdensities (computed in \citealt{Cucciati2017}). By this, we checked the loci of galaxies that most reside in less-dense environments and overdense ones. We found subtle difference in this relation, galaxies in over-dense environments were above the fitted IRX-$\beta$ relation (Equation \ref{eq-fit}), and galaxies in under-dense regions were below this fit. However, given the large error bars, this difference was not robust. Similarly we tested the attenuation in the FUV bands of our galaxies depending on the overdensity of their environments in Figure \ref{fig:env_fuv}. We conclude that the environment does not affect dust attenuation of our sample of star-forming galaxies at intermediate redshift, despite the known effect the environment has on main galaxy properties such as the SFRs \citep{Peng2010}. Similar results were found by \citet{Shivaei2020} around the cosmic noon.



\begin{acknowledgements}
M.H. acknowledges the support by the National Science Centre, Poland (UMO-2022/45/N/ST9/01336). K.M, M.H. and J. have been supported by the National Science Centre (UMO-2018/30/E/ST9/00082). A.N. acknowledges support from the Narodowe Centrum Nauki (UMO-2020/38/E/ST9/00077). A.P., F.P., J.K.  were supported by the Polish National Science Centre grant (UMO-2018/30/M/ST9/00757). This research was supported by Polish Ministry of Science and Higher Education grant DIR/WK/2018/12.
M.H. thanks M\'ed\'eric Boquien and Stephane Arnouts for the discussion.

\end{acknowledgements}

\bibliographystyle{aa}
\bibliography{aanda.bib}

\begin{appendix}

\section{Comparison of SED-computed SFRs with tracers}\label{appendixa}
\begin{figure}[hb!]
    \centering
    \includegraphics[width=1\columnwidth]{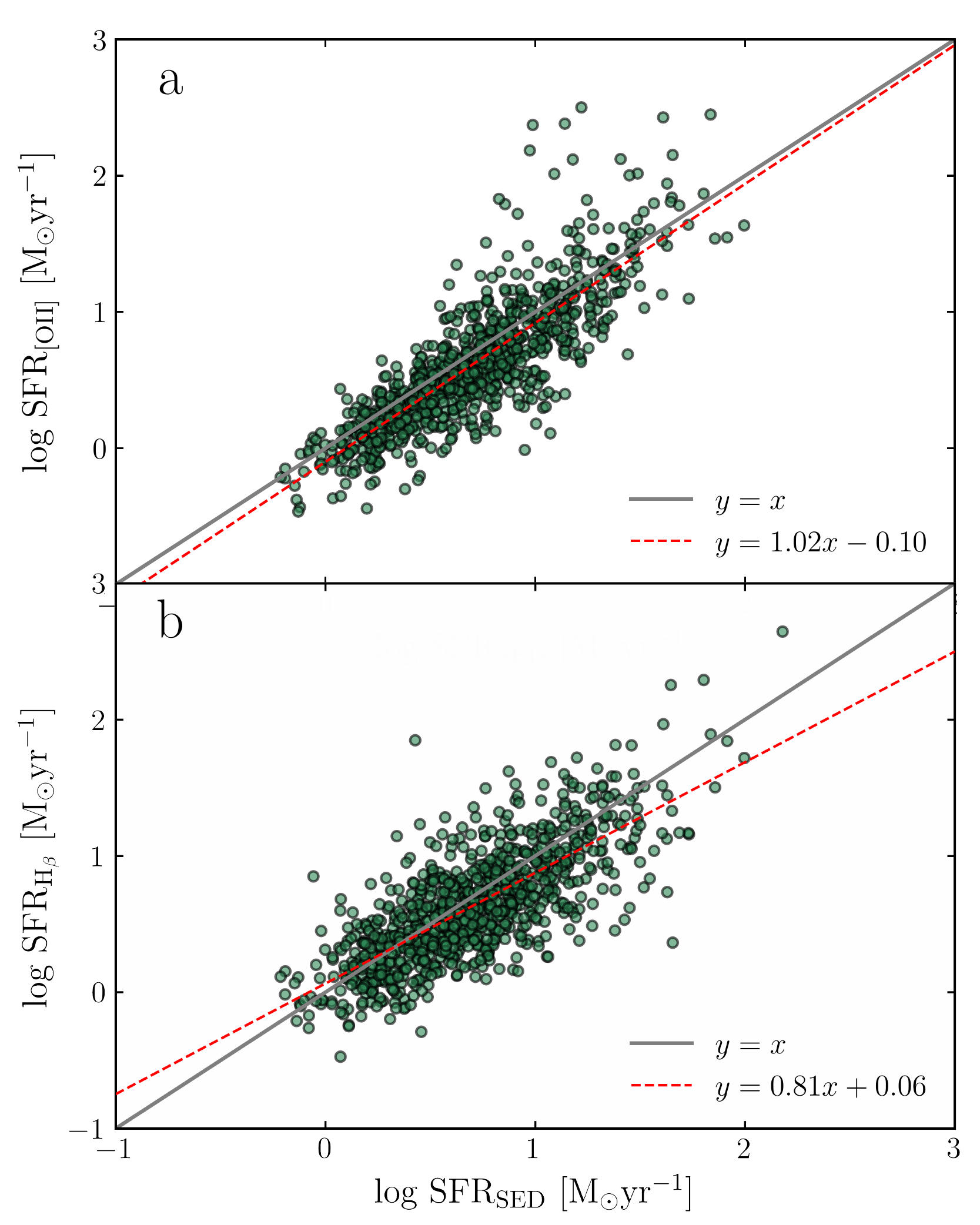}
    \caption{Comparison between star formation rates derived with (a):$\ion{O}{ii}$, (b):H$_{\beta}$ with those derived with SED fitting. The red dashed line is the best fit of the distribution.}
    \label{fig:sfrs}
\end{figure}

We show in Figure \ref{fig:sfrs} the comparisons between the SFRs computed using \texttt{CIGALE} and those computed using the emission lines H$\beta$ and $\ion{O}{ii}$. The correlation between all these estimators demonstrate the effectiveness of the derived SFRs using panchromatic SED fitting.\\
To compute the SFRs from $\ion{O}{ii}$, we used the method described in \citet{Kennicutt98}: \[log(\mathrm{SFR}(\ion{O}{ii})\ [\mathrm{M}_{\odot}\mathrm{yr}^{-1}]) = log(\mathrm{L}(\ion{O}{ii})\ [\mathrm{erg\ s^{-1}}]) - 41.20.\]
To compute the SFRs from H$_{\beta}$, we assume H$_{\alpha}$/H$_{\beta}$=2.86 \citep{Figueira2022}. Then, we computed the SFRs using the \citet{Kennicutt98} relation for H$_{\alpha}$ which is given by: \[log(\mathrm{SFR}(H_{\alpha})\ [\mathrm{M}_{\odot}\mathrm{yr}^{-1}]) = log(\mathrm{L}(H_{\alpha})\ [\mathrm{erg\ s^{-1}}]) - 41.10\]
Both ${L}(\ion{O}{ii})$ and H$_{\beta}$ are corrected for attenuation as described in Section \ref{metal}.

\section{Mass-complete sub-sample}\label{appendixb}
We also derive the IRX-$\beta$ relation for our mass-complete sub-sample. To achieve the mass-complete sub-sample, we select galaxies for which log(M$_{\star}$) $\geq$ 10.18 M$_{\odot}$ for $z \leq 0.6$, and log(M$_{\star}$) $\geq$ 10.47 M$_{\odot}$ for $z \leq 0.8$, following \citet{Davidzon2016} selections for VIPERS.\\

\begin{figure}
    \centering
    \includegraphics[width=1\columnwidth]{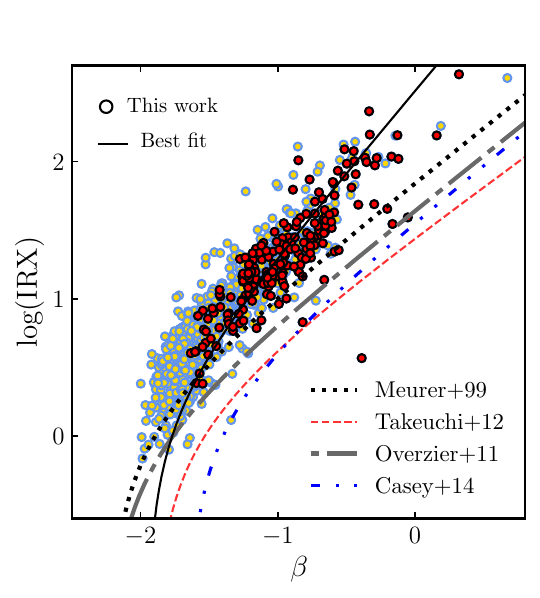}
    \caption{IRX-$\beta$ scatter of our sample. The original \citet[][]{Meurer99} fit is shown in dotted line. We also compare with fits from \citet{Overzier2011, Takeuchi2012, Casey2014}, denoted with grey dash-dotted, red dashed, and dash-double dotted lines, respectively. The solid black line shows Equation \ref{eq-mass_complete}, describing our mass-complete sample. Red dots denote the mass-complete sub-sample in this work.}
    \label{fig:irx-beta-masscomplete}
\end{figure}

These selections result in only 258 galaxies from our initial sample. The IRX-$\beta$ fit for this sub-sample is then given by:
\begin{equation}
   \mathrm{IRX} = \mathrm{log}[(10^{2.98\beta+1.95}-1)/0.64].
    \label{eq-mass_complete}
\end{equation}
We show the fit for the mass-complete sub-sample in Figure \ref{fig:irx-beta-masscomplete}. The mass-complete sample is statistically less important than our full sample, and it lacks data in the lower IRX and $\beta$ values, therefore it must be considered with caution.
\end{appendix}

\end{document}